%% file: Enhancing_utility.tex
\documentclass{article}
\usepackage{spconf,amsmath,graphicx,amssymb,mathtools, nccmath,mathrsfs,subfigure}
\usepackage[lined,linesnumbered,ruled,commentsnumbered]{algorithm2e} 
\usepackage[belowskip=-10pt,aboveskip=5pt]{caption}
\usepackage{url}
\usepackage{ifthen}
\usepackage{cite}
\usepackage{cancel}
\usepackage{enumerate}
\usepackage[normalem]{ulem}
\usepackage{tikz}
\usepackage{pgfplots}
\pgfplotsset{colormap/jet}

\usepackage{blkarray}
\usepackage{diagbox}
\usepackage{blindtext}

\newenvironment{Ni}[1]{\textcolor{black}{#1}}

\newenvironment{Zr}[1]{\textcolor{black}{#1}}

\usetikzlibrary{arrows,shapes,backgrounds,plotmarks,positioning}

\pgfplotsset{compat=newest}                         
\pgfplotsset{plot coordinates/math parser=false}
\newlength\figureheight
\newlength\figurewidth

\DeclareMathOperator*{\argmax}{arg\,max}
\DeclareMathOperator*{\argmin}{arg\,min}

\newtheorem{definition}{Definition}

\newtheorem{example}{Example}

\newcommand{\Sen}{\mathcal{S}}
\newcommand{\X}{\mathcal{X}}
\newcommand{\Y}{\mathcal{Y}}

\newcommand{\Psx}{p(s,x)}

\newcommand{\Pygx}{p(y|x)}

\newcommand{\Set}[1]{\{#1\}}

\newcommand{\eps}{\varepsilon}
\newcommand{\Xeps}{\X_\varepsilon}

\newcommand{\Xepsc}{\X_\varepsilon^c}

\newcommand{\Xell}{\X_{\varepsilon}}
\newcommand{\Xh}{\X_{\varepsilon}^{c}}
\newcommand{\Xhi}[1]{\X_{#1}}

\newcommand{\Rhi}[1]{R_{#1}(y)}

\title{Enhancing Utility in the Watchdog Privacy Mechanism}

\name{Mohammad Amin Zarrabian$^{\star}$ \qquad Ni Ding$^{\ast}$ \qquad Parastoo Sadeghi$^{\dagger}$ \qquad Thierry Rakotoarivelo$^{\ddagger}$\thanks{This work is partly supported by the Data61 CRP: IT-PPUB and by the ARC Future Fellowship FT190100429.}}

\address{$^{\star}$  College of Engineering and Computer Science, Australian National University, Canberra, Australia.\\
	$^{\ast}$School of Computing and Information Systems, University of Melbourne, Melbourne, Australia.\\
	$^{\dagger}$School of Engineering and Information Technology, University of New South Wales, Canberra, Australia.\\
	$^{\ddagger}$ Data61, Commonwealth Scientific and Industrial Research Organisation, Eveleigh, Australia.\\
\small{mohammad.zarrabian@anu.edu.au, ni.ding@unimelb.edu.au, p.sadeghi@unsw.edu.au, thierry.rakotoarivelo@data61.csiro.au.}}

\begin{document}
	%
	\maketitle
	\begin{abstract}
		
		This paper is concerned with enhancing data utility in the privacy watchdog method for attaining information-theoretic privacy. For a specific privacy constraint, the watchdog method filters out the high-risk data symbols through applying a uniform data regulation scheme, e.g., merging \emph{all} high-risk symbols together. While this method entirely trades the symbols resolution off for privacy, we show that the data utility can be greatly improved by partitioning the high-risk symbols set and individually privatizing each subset. We further propose an agglomerative merging algorithm that finds a suitable partition of high-risk symbols: it starts with a singleton high-risk symbol, which is iteratively fused with others until the resulting subsets are private.~Numerical simulations demonstrate the efficacy of this algorithm in privately achieving higher utilities in the watchdog scheme.
		
	\end{abstract}


	\section{Introduction}
		\label{sec:intro}
		 Industries and governments are increasingly sharing data to unlock economic and societal benefits through advances in data analytics and machine learning.
	 	However, such data also contains sensitive information about individuals or businesses, which makes the privacy regulators, users, and data providers concerned about the leakage of confidential information, either explicitly or implicitly.
		In signal processing and information theory, data privacy is underpinned in terms of a measure called \emph{information lift}\cite{PvsInfer2012,Watchdog2019}.

		To evaluate how much private data \Ni{$X$} is \Zr{informative} about the confidential \Zr{data} \Ni{$S$}, the lift measures the change in the posterior belief $p(s|x)$ from the prior belief $p(s)$ \Ni{for each instance of $s$ and $x$} by
		\begin{equation}\label{equ:lift}
			l(s,x)=\frac{p(s|x)}{p(s)}.
		\end{equation}
		It is clear that a small lift indicates limited private information gain by the adversary and therefore the more private \Ni{$X$} is.
		The lift is the elementary measure in almost all information leakage measures, e.g., mutual information \cite{PvsInfer2012,PF2014}, Sibson mutual information  \cite{Sibson1969,Verdu2015AlphaMI, alphading2021}, $\alpha$-leakage \cite{Liao2019Alpha} and local differential privacy \cite{LDP2013MiniMax,LDP}: as proved in \cite{Watchdog2019}, if lift is bounded, all these leakage measures are also bounded.

		\begin{table}[t]
			\centering 	
			\begin{tabular}{|c|c|c|}
				\hline
				$\eps=1$ & utility &  privacy leakage \\
				\hline
				complete merging & 0.5913 & 0.8037 \\
				\hline
				two-subset merging & 0.7335 & 0.8488 \\
				\hline
			\end{tabular}	
			\caption{\label{tab1} An example of how subset merging can enhance utility in the watchdog mechanism.}
		\end{table}		
	
		The \Ni{existing} approach to attain lift-based privacy is the watchdog method proposed in \cite{Watchdog2019}. For a specific privacy constraint, i.e., a threshold $\epsilon$ on the lift, the watchdog method filters out and privatizes the high-risk symbols of \Ni{$X$}. The authors in \cite{Watchdog2019} adopted a uniform approach: merging all high-risk symbols together into a `super' symbol, which is proved in \cite{Sadeghi2020ITW, alphading2021} to be the optimal privatization scheme in attaining data privacy.
\Ni{However, this uniform approach neglects an important issue in data privacy: to achieve the benefits of data sharing, the privatized data should provide a satisfactory level of usefulness in $X$.\footnote{The data utility is usually quantified by average performance measures such as mutual information \cite{PF2014}, $f$-divergence \cite{f-diveregence-utiltiy} and Hamming distortion \cite{LDP2014Lalitha}. We use mutual information in this paper.} Despite the relaxation attempts in \cite{Sadeghi2020ITW, alphading2021}, the \emph{complete merging} method minimizes the resolution of high-risk symbols and significantly deteriorates data utility, which is at odds with the purpose of data sharing.}
	
		%
	%

		In fact, even a small alteration can greatly enhance the data utility. In Table~\ref{tab1}, we arbitrarily cut the high-risk symbol set (of size 7) into two subsets, each of which is then privatized individually. The utility (measured by mutual information) is increased significantly without sacrificing too much data privacy, which remains below the design constraint $\epsilon$. This not only shows that the complete merging approach is an `overkill' in terms of data utility, but also suggests a partitioned privatization approach.

		In this paper, we propose the subset merging method for watchdog mechanism to enhance the data utility.
		Finding the best partition of high-risk symbols set to achieve optimal utility is generally a combinatorial problem.
		Accordingly, we propose a heuristic greedy algorithm to find good subsets to merge, which guarantees data privacy. To do so, this greedy algorithm tries to search the finest partition of the \Ni{original} high-risk symbols set that ensures the \Ni{resulting} lift \Ni{of the whole dataset} does not exceed $\epsilon$.
		It starts with the singleton partition of high-risk symbols (highest resolution) and iteratively merges symbols until lift values of the resulting subsets are all below $\epsilon$. Numerical simulations show that our proposed algorithm enhances utility significantly and maintains the privacy leakage constraint.

	\section{System Model}
	

\Ni{Denote random variables $S$ and $X$ the sensitive and public data, respectively. The joint distribution $p(s,x)$ describes the statistical correlation between $S$ and $X$. To protect the privacy of $S$, we sanitize $X$ to $Y$ by the transition probability $p(y|x)$. Here, for each $x,y$, $p(y|x)=p(y|x,s), \forall s$ and therefore the Markov chain $S \rightarrow X \rightarrow Y$ is formed. }
		%

\Ni{The watchdog method is based on the logarithm of the lift measure 
\[ i(s;x)=\log{l(s; x)}. \]
For each $x \in \X$, denote the maximum symbol-wise information leakage by $\max_{s \in \Sen}|i(s,x)|$, where
		\begin{equation}
			\omega(x)=\max_{s \in \mathcal{S}}|i(s,x)|.
		\end{equation}
Applying an upper bound $\epsilon$ to $\omega(x)$ for all symbols $x \in \X$, the whole alphabet $\X$ is divided into two subsets:  the low-risk subset is given by
		$\Xeps \triangleq \{x\in\X: \omega(x)\leq\eps\}$
that is safe to publish, and the high-risk symbol set
$$ \Xh=\X\setminus\Xell\triangleq \{x\in\X: \omega(x)>\eps\} $$
that requires some treatment before the data publishing. }

\Ni{The authors in \cite{Sadeghi2020ITW,alphading2021} adopt a uniform randomization scheme 
\begin{equation} \label{eq:origina watchdog}
			\Pygx = \begin{cases}
				1_{\{x=y\}} & x , y \in \Xeps,\\
				R(y)    	& x , y \in\Xh,\\
				0 		  	& \text{otherwise},
			\end{cases}
		\end{equation}
where $R(y)$ is \emph{complete merging} solution, e.g., where there is only one super symbol $y^* \in \Xh$ such that $R(y^*) = 1$ for all $x \in \Xepsc$ and $R(y)=0$ otherwise.
}

\begin{algorithm}[t]
	\textbf{Input}: $\X,\eps, \Psx$\\
	\textbf{Output}: $\mathcal{P}_{\Xh}=\{\Xhi{1},\Xhi{2},\cdots\,\X_p\}$ \\
	\textbf{Initialize}: Obtain $\{\Xeps,\Xepsc\}$, $\X_{Q} \leftarrow \Xepsc$, and $p=1$ 
	\While{$|\X_{Q}|>0$}
	{
		$\X_p=\{\displaystyle\argmax_{x\in \X_{Q}}{\omega(x)}\}$, and
		$\X_{Q} \leftarrow\X_{Q}\setminus \X_p$;
		
		\While{$\omega({\X_p})>\eps$ \& $|\X_{Q}|>0$}
		{
			$x^{*}=\displaystyle\argmin_{x \in \X_{Q}}\omega(\X_p \cup \{x\})$\\
			$\X_p \leftarrow \X_p \cup \{x^{*}\}$ and $\X_{Q}\leftarrow \X_{Q}\setminus\{x^{*}\}$;
		}
		$\mathcal{P}_{\X_Q}=\{\X_1,\X_2,\cdots,\X_p\}$, and  $p \leftarrow p+1$ 	
	}	
	
	\While {$\omega(\X_p)>\eps$ \& $|\mathcal{P}_{\X_Q}|>1$ }{
		$\X_{k}=\displaystyle\argmin_{1<i<p}\omega(\X_p\cup\X_i)$;\\
		$\X_p \leftarrow \X_p\cup\X_k$; for $k+1\leq i \leq p$ update the index of $\X_i$'s to $\X_{i-1}$\\
		$\mathcal{P}_{\X_Q}=\{\X_1,\X_2,\cdots,\X_p\}$}
	
	\caption{Make a refinement of $\Xepsc$ } \label{algorithm1}	
\end{algorithm}

After randomization, the log-lift is given by $i(s,y)=\log\frac{p(y|s)}{p(y)}$ where $p(y|s)=\sum_{ x \in \X}p(y|x)p(x|s)$ due to the Markov property and $p(y)=\sum_{ x \in \X}p(y|x)p(x)$.
		We can extend the notion of the log-lift, and $\omega(x)$ from a single $x \in \X$ to a subset $\X_{Q}\subseteq \X$ \cite{Sadeghi2020ITW}:
		\begin{equation}
			i(s,\X_{Q})=\log \frac{p(\X_{Q}|s)}{p(\X_{Q})}, \quad \omega(\X_{Q})=\max_{s \in \Sen}|i(s,\X_{Q})|,
		\end{equation}
		where $p(\X_{Q}|s)=\sum_{ x \in \X_Q}p(x|s)$ and $p(\X_{Q})=\sum_{ x \in \X_Q}p(x)$.
	%
	
\Ni{Applying $p(y|x)$, }the value of $\max_{y \in \Y}\omega(y)$ as the upper bound on privacy leakage after randomization is given by \cite{Sadeghi2020ITW}
		\begin{equation}
			\max_{y \in \Y}\omega(y)=\max\{\max_{y\in \Xeps}\omega(y),\omega(\Xepsc)\}.
		\end{equation}
		It is obvious that $\max_{y\in \Xeps}\omega(y) \leq \eps$, so it attains the privacy constraint. However, the value of $ \omega(\Xepsc)$ is variable and depends on the joint probability distribution.
		
		\begin{example}\label{example 1}
			Let $\X=\{x_1,x_2,\cdots,x_5\}$, $\Sen=\{s_1,s_2,s_3\}$, and $\eps=0.8$. We randomly generate a joint distribution $p(s,x)$ and the resulting maximum symbol-wise leakage and utility are $\omega(x)=[1.3515, 1.6458, 0.9295,  0.8161,  0.2608]$ and $H(X)=1.6034$, respectively.
			For the given $\eps$, the low-risk and high-risk subsets are given by $\Xeps=\{x_5\}$ and $\Xepsc=\{x_1,x_2,x_3,x_4\}$, respectively. After randomization, assume high-risk symbols are mapped to $y^{*}$ where $y^{*}=y_1=x_1$ and $y_2=x_5$, then the maximum symbol-wise leakage and utility are given by $\omega(y)=[0.0627, 0.2608]$ and $I(X;Y)=0.5269$, respectively.
		\end{example}
		
		In Example \ref{example 1},  the leakage in the high-risk subset after randomization is $\omega(\Xepsc)=0.0627$, which is \Zr{an order of magnitude} smaller than the original privacy constraint $\eps=0.8$, based on which $\Xepsc$ was obtained. Although this small leakage guarantees a very high level of privacy, \Zr{it damages utility drastically, the utility decreases from $H(X)=1.6034$ to  $I(X;Y)=0.5269$.} On the other hand, when a threshold $\eps$ is given as the privacy constraint, it is acceptable to just keep the privacy leakage less than $\eps$, even if it becomes very close to this threshold. Thus, in the next section, we propose a subset merging method to enhance utility where the privacy leakage increases, but remains under $\eps$ to the extent possible.

	\section{Enhancing Utility}
	
		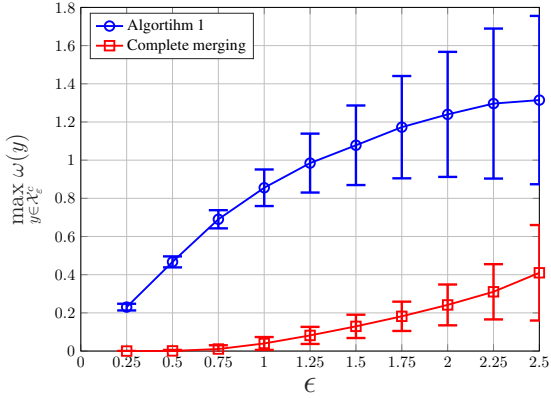
\begin{figure}[t]
		\centering
		\scalebox{0.6}{\input{figures/Privacy.tex}}
		\caption{Privacy leakage for different values of $\eps$: The mean values of $\max_{y\in \Xepsc}\omega(y)$ are shown with standard deviation. Algorithm \ref{algorithm1} increases privacy leakage in $\Xepsc$ in comparison with complete merging, however, in all cases it is still below the constraint $\eps$.}\smallskip
		\label{fig:Privacy}
	\end{figure}
	
		In this section, we introduce an approach to enhance utility while maintaining a set privacy constraint.
		We measure utility by mutual information which for the bi-partition $\{\Xeps,\Xepsc\}$ and complete merging randomization $\Xepsc$ is given by \cite{Sadeghi2020ITW}
		\begin{equation}\label{eq:muinfo}
			I(X;Y) =H(X)+ \sum_{x \in \Xh} p(x) \log \frac{p(x)}{p\left(\Xh\right)}.
		\end{equation}
		Clearly, the utility depends on $p(x)$ for $x\in \Xepsc$ and the overall $p(\Xepsc)$. Our proposed approach to enhance data utility is through increasing data resolution. That is, we propose to randomize subsets of $\Xepsc$ separately rather than the complete merging of the whole set $\Xepsc$.
	
		Let $[p]=\{1,2,\cdots,p\}$ and consider a bi-partition $\Set{\Xeps,\Xepsc}$, a further partitioning of elements in $\Xepsc$ denoted by $\mathcal{P}_{\Xepsc}=\{\Xhi{1},\cdots,\Xhi{p}\}$, and complete merging randomizations $\Rhi{i}, i \in [p]$  where $\sum_{y\in\Xhi{i}}\Rhi{i}=1$.
		In other words, we partition $\Xh$ to subsets $\Xhi{i}$, so $\Xh=\cup_{i=1}^{p}\Xhi{i}$ and each subset $\Xhi{i}$ is randomized by the corresponding randomization $\Rhi{i}$. Note that for $y \in \X_i$ we have $p(y)=\sum_{x \in \X_i}p(y|x)p(x)=p(\X_i)\Rhi{i}$.
		The resulting mutual information $I(X;Y)$ is
		\begin{equation}\label{eq:muifo-subset}
			I(X;Y) =H(X)+\sum_{i=1}^{p}  \sum_{ x \in \Xhi{i}} p(x) \log \frac{p(x)}{p\left(\Xhi{i}\right)}.
		\end{equation}
		Then the normalized mutual information-loss for partition $\mathcal{P}_{\Xepsc}$ on $\Xh$ is defined as
		\begin{align}
			\text{NMIL}(\mathcal{P}_{\Xh})&=\frac{H(X)-I(X;Y)}{H(X)}.
		\end{align}
		Since $p(\Xhi{i})\leq p(\Xepsc)$ for $i \in [p]$ the data resolution increases for each $x \in \Xhi{i}$ and this can results in a larger mutual information and hence a lower utility loss. The following definition and derivations formalize this observation.
		\begin{definition}
			Assume two partitions $\mathcal{P}_{\Xh}=\{\Xhi{1},\cdots,\Xhi{p}\}$ and $\mathcal{P}_{\Xh}^{'}=\{\Xhi{1}^{'},\cdots,\Xhi{p'}^{'}\}$. We say  $\mathcal{P}_{\Xepsc}^{'}$  is a refinement of $\mathcal{P}_{\Xepsc}$ and $\mathcal{P}_{\Xh}$  is an aggregation of $\mathcal{P}_{\Xh}^{'}$ \cite{Liuindexcoding},  if for every $i \in [p],$ $\Xhi{i}=\cup_{j \in J_i}\Xhi{j}^{'}$ where $J_i\subseteq [p']$, and we have $p(\Xhi{i})=\sum_{j\in J_i}{p(\Xhi{j}^{'})}$.
		\end{definition}
		If $\mathcal{P}_{\Xh}^{'}$ is a refinement of $\mathcal{P}_{\Xh}$ and  $\mathcal{P}_{\Xh}$ is an aggregation of $\mathcal{P}_{\Xh}^{'}$  then $\text{NMIL}(\mathcal{P}_{\Xh}^{'})\leq\text{NMIL}(\mathcal{P}_{\Xh})$. This is because
		\begin{align}
			&H(X)\times\text{NMIL}(\mathcal{P}_{\Xh})=\sum_{i=1}^{p} \sum_{x\in\Xhi{i}}p(x)\log\frac{p(\Xhi{i})}{p(x)}\\ \nonumber
			&\geq\sum_{i=1}^{p'} \sum_{x\in\Xhi{j}}p(x)\log\frac{p(\Xhi{j})}{p(x)}=H(X)\times\text{NMIL}(\mathcal{P'}_{\Xh}).
		\end{align}
	
		Generally, finding an optimal partition $\mathcal{P}_{\Xh}$ that maximizes utility while maintaining the privacy constraint is combinatorial since it depends on the high-risk symbol probabilities $p(x)$, $x \in \Xepsc$ and the joint probability distribution. However, we can use $\eps$ as a stop criteria to make a heuristic agglomerative algorithm for obtaining a refinement of $\Xepsc$ to enhance utility as much as possible.

	\subsection{A greedy algorithm to refine the high-risk subset $\Xepsc$}

\Ni{Knowing that the aggregation of the partition of $\Xepsc$ reduces the symbol resolution and data utility, we propose a heuristic greedy algorithm that determines the most refined partition of $\Xepsc$ that satisfies the data privacy constraint specified by $\epsilon$. This is a bottom-up algorithm, which bootstraps from the most refined (singleton-element) partition of $\Xepsc$. This starting point provides the highest resolution/utility but results in a lowest data privacy level. We let the subsets in the singleton partition merge with each other to reduce the log-lift measure \Zr{$\omega(\Xhi{i}), \Xhi{i} \in \mathcal{P}_{\Xepsc}$}  until the log-lift of all subsets is reduced below $\epsilon$.
To achieve a low complexity, but effective procedure that results in a finer partition of $\Xepsc$,  we implement the subset merging in order. A good candidate to begin with is the most challenging symbol with the highest log-lift leakage.
}

The pseudo-code of our method is shown in Algorithm \ref{algorithm1}. To find each subset $\X_i \subseteq \Xepsc$, we start with $\argmax_{x\in \Xepsc}\omega(x)$ as the symbol with the highest risk in $\Xepsc$ (line 4) and then to make the leakage of the subset $\omega(\X_i)$ less than $\eps$ we merge another symbol to it that minimizes $\omega(\X_i)$ in each iteration (lines 5-8). Each symbol that is added to $\X_i$ is removed from $\Xepsc$ (line 7). When a subset $\Xhi{i}$ is made, we add it to the partition set $\mathcal{P}_{\Xepsc}$ and repeat the same process for the remaining $x \in \Xh$.
		After making all subsets, there is a possibility that for the last subset $\Xhi{p}$ the leakage is greater than $\eps$. That is, $\omega(\X_p)>\eps$.  If this happens, while $\omega(\X_p)>\eps$ we make an agglomerate $\X_p$ by merging a subset to it that minimizes $\omega(\X_p)$ (lines 11-15). A complete merging is a special output of our algorithm if no better finer partition can be found that maintains the privacy.


		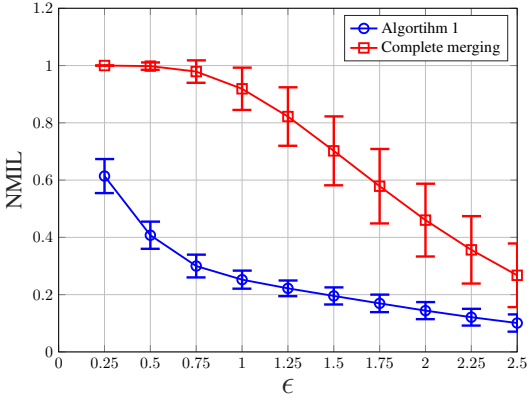
\begin{figure}[t]
			\centering
			\scalebox{0.6}{\input{figures/NMIL.tex}}
			\caption{Normalized Utility loss (NMIL) for different values of $\eps$: The mean values of NMIL are shown with standard deviation. Algorithm \ref{algorithm1} reduces NMIL in comparison with complete merging.}\smallskip
			\label{fig:NMIL}
		\end{figure}

	\section{Experiments}

		To make a comparison between Algorithm \ref{algorithm1} and complete merging, we randomly generated 1000 joint distributions $p(s,x)$ where $|\X|=20$ and $|\mathcal{S}|=13$. For each distribution, after randomization we obtained maximum privacy leakage of high-risk symbols $\max_{y\in \Xepsc}\omega(y)$, maximum overall privacy leakage $\max_{y \in \Y}\omega(y)$, and the utility loss NMIL under Algorithm \ref{algorithm1} and complete merging for different values of $\eps\in\{0.25,0.5,0.75,\cdots,2.25,2.5\}$. Then for each $\eps$, we derived the mean value and standard deviation of $\max_{y\in \Xepsc}\omega(y)$, $\max_{y}\omega(y)$, and NMIL across these 1000 observations.

		In Fig. \ref{fig:Privacy} the mean value of  $\max_{y\in \Xepsc}\omega(y)$ is depicted for each $\eps$, as well as its standard deviation (shown as tolerance bars).
		As expected, complete merging makes a strong  guarantee on privacy leakage and keeps both mean and standard deviation much less than $\eps$ in all cases.
		In contrast, algorithm \ref{algorithm1} increases the privacy leakage and lets it be closer to $\eps$ compared to complete merging, but crucially it still keeps the mean value and the corresponding deviation less than $\eps$ in all cases.
		\Zr{As the value of $\eps$ increases, the standard deviation is also increased. This is because when $\eps$ increases, the  privacy constraint is less strict and consequently the size of $\Xepsc$ decreases.
		As a result, the sample size to calculate the mean value of privacy leakage reduces, which causes a larger deviation.}
	
		Fig. \ref{fig:NMIL} shows the normalized mutual information loss under Algorithm $\ref{algorithm1}$ and complete merging. It demonstrates that Algorithm \ref{algorithm1} enhances utility substantially for each value of $\eps$.
		
		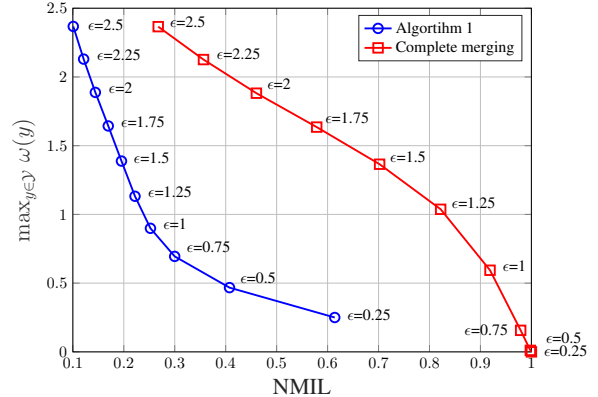
\begin{figure}[t]
			\centering
			\scalebox{0.6}{\input{figures/PUT.tex}}
			\caption{Privacy Utility trade-off: For each value of $\eps$, algorithm \ref{algorithm1} reduces NMIL substantially while maintain privacy leakage below $\eps$.}\smallskip
			\label{fig:PUT}
		\end{figure}
	
		Finally, to have a clear privacy-utility trade-off (PUT) comparison, we present PUT curves for both Algorithm \ref{algorithm1} and complete merging in Fig. \ref{fig:PUT}.
		For each $\eps$, the mean value of overall privacy leakage $\max_{y \in \Y}\omega(y)$ is shown versus the corresponding mean value of NMIL.
		Clearly, Algorithm \ref{algorithm1} enhances utility significantly while satisfying the privacy leakage constraint. For a very strict constraint on privacy ($\eps\leq 0.5$), complete merging results in perfect privacy with total utility loss where $\text{NMIL}=1$. However, Algorithm $\ref{algorithm1}$ keeps the average utility loss less than $1$.

	\section{Conclusion}
		In this paper, we introduced a method to enhance utility in the watchdog privacy mechanism. We showed that it is possible to maintain the privacy constraint and improve utility. In our approach, instead of randomizing the whole high-risk partition, we randomized subsets of high-risk symbols separately. Then we proposed a heuristic greedy algorithm to find subsets of high-risk elements and at the same time keep the leakage of each subset less than the privacy constraint, $\eps$. The simulation results showed substantial utility enhancement and preservation of the privacy constraint.
	
		In future research, it would be interesting to consider how to develop more efficient subset merging algorithms. Investigation of effective parameters like subset size and different privacy and utility measures could be helpful. It would be also beneficial to apply the proposed algorithm to real world data sets and measure actual practical utility. It would be also beneficial to apply the proposed algorithm to real world datasets and measure the obtained utility.

	\bibliographystyle{IEEEbib}
	\bibliography{BIB}
	
\end{document}

%% file: figures/Privacy.tex
%
%
\begin{tikzpicture}

\begin{axis}[%
width=4in,
height=3in,
scale only axis,
xmin=0,
xmax=2.5,
xtick={0, 0.25,  0.5, 0.75,    1, 1.25,  1.5, 1.75,    2, 2.25,  2.5},
xlabel style={font=\color{white!15!black}},
xlabel={\huge $\epsilon$},
ymin=0,
ymax=1.8,
ylabel style={font=\color{white!15!black}},
ylabel={\Large $\displaystyle \max_{y\in \Xepsc} \hspace{2pt}\omega(y)$ },
axis background/.style={fill=white},
title style={font=\bfseries},
xmajorgrids,
ymajorgrids,
legend style={at={(0.02,0.85)}, anchor=south west, legend cell align=left, align=left, draw=white!15!black}
]
\addplot [color=blue, line width=1.2pt, mark=o, mark options={solid, blue},  mark size=3.0pt]
 plot [error bars/.cd, y dir=both, y explicit, error bar style={line width=1.5pt}, error mark options={line width=1.5pt, mark size=6.0pt, rotate=90}]
 table[row sep=crcr, y error plus index=2, y error minus index=3]{%
0.25	0.230267534361982	0.0174566889722763	0.0174566889722763\\
0.5	0.467106283323074	0.0288360602215737	0.0288360602215737\\
0.75	0.690323872117771	0.0475689545013867	0.0475689545013867\\
1	0.855380705152294	0.0957796797771706	0.0957796797771706\\
1.25	0.984561443958781	0.154307595145896	0.154307595145896\\
1.5	1.07793802898244	0.208353318073746	0.208353318073746\\
1.75	1.17275716716679	0.267920769427333	0.267920769427333\\
2	1.23964899490579	0.327497214947916	0.327497214947916\\
2.25	1.29636905148116	0.392847630133167	0.392847630133167\\
2.5	1.31463399229699	0.440834975845881	0.440834975845881\\
};
\addlegendentry{Algortihm 1}

\addplot [color=red, solid, line width=1.2pt, mark=square, mark options={solid, red}, mark size=3.0pt]
 plot [error bars/.cd, y dir=both, y explicit, error bar style={line width=1.5pt}, error mark options={line width=1.5pt, mark size=6.0pt, rotate=90}]
 table[row sep=crcr, y error plus index=2, y error minus index=3]{%
0.25	1.43773881688958e-16	1.45845477780384e-16	1.45845477780384e-16\\
0.5	0.000793822070058047	0.00508549126642078	0.00508549126642078\\
0.75	0.0106192795257784	0.0198395886473911	0.0198395886473911\\
1	0.0397823673342588	0.0336370942326327	0.0336370942326327\\
1.25	0.0819102353101155	0.0447317890784551	0.0447317890784551\\
1.5	0.129353868717252	0.0610726438414224	0.0610726438414224\\
1.75	0.18199669375157	0.0767069866264454	0.0767069866264454\\
2	0.241692058051209	0.106917726248026	0.106917726248026\\
2.25	0.310460923183735	0.144696311572706	0.144696311572706\\
2.5	0.409986902499499	0.250101942447674	0.250101942447674\\
};
\addlegendentry{Complete merging}

\end{axis}
\end{tikzpicture}%

%% file: figures/NMIL.tex
%
%
\begin{tikzpicture}

\begin{axis}[%
width=4in,
height=3in,
scale only axis,
xmin=0,
xmax=2.5,
xtick={0, 0.25,  0.5, 0.75,    1, 1.25,  1.5, 1.75,    2, 2.25,  2.5},
xlabel style={font=\color{white!15!black}},
xlabel={\huge $\epsilon$},
ymin=0,
ymax=1.2,
ylabel style={font=\color{white!15!black}},
ylabel={\Large NMIL},
axis background/.style={fill=white},
title style={font=\bfseries},
xmajorgrids,
ymajorgrids,
legend style={legend cell align=left, align=left, draw=white!15!black}
]
\addplot [color=blue, line width=1.2pt, mark=o, mark options={solid, blue},  mark size=3.0pt]
 plot [error bars/.cd, y dir=both, y explicit, error bar style={line width=1.5pt}, error mark options={line width=1.5pt, mark size=6.0pt, rotate=90}]
 table[row sep=crcr, y error plus index=2, y error minus index=3]{%
0.25	0.614055979843306	0.0594256087546705	0.0594256087546705\\
0.5	0.40749218006555	0.0474282750176185	0.0474282750176185\\
0.75	0.299882004727968	0.0397535093769401	0.0397535093769401\\
1	0.252336877359735	0.0315371977421828	0.0315371977421828\\
1.25	0.222001506768911	0.0273523577229689	0.0273523577229689\\
1.5	0.195460632929228	0.0298535467965817	0.0298535467965817\\
1.75	0.169494509817045	0.0305768001447222	0.0305768001447222\\
2	0.144196061764307	0.0298069287332999	0.0298069287332999\\
2.25	0.121211717113631	0.0292487964044727	0.0292487964044727\\
2.5	0.100839132874635	0.0301093886868998	0.0301093886868998\\
};
\addlegendentry{Algortihm 1}

\addplot [color=red, solid, line width=1.2pt, mark=square, mark options={solid, red}, mark size=3.0pt]
 plot [error bars/.cd, y dir=both, y explicit, error bar style={line width=1.5pt}, error mark options={line width=1.5pt, mark size=6.0pt, rotate=90}]
 table[row sep=crcr, y error plus index=2, y error minus index=3]{%
0.25	1	8.10944420711344e-17	8.10944420711344e-17\\
0.5	0.997978869972794	0.0128045705411678	0.0128045705411678\\
0.75	0.9789580229695	0.0391652468835089	0.0391652468835089\\
1	0.918817040609376	0.0738118014365136	0.0738118014365136\\
1.25	0.821946801040326	0.10211723872564	0.10211723872564\\
1.5	0.70211502622707	0.120325376837655	0.120325376837655\\
1.75	0.578666240060208	0.129990342551709	0.129990342551709\\
2	0.460278819984222	0.127099030383038	0.127099030383038\\
2.25	0.356422013842445	0.117701119556063	0.117701119556063\\
2.5	0.267552114870886	0.111089847729876	0.111089847729876\\
};
\addlegendentry{Complete merging}

\end{axis}

\end{tikzpicture}%

%% file: figures/PUT.tex
%
%
\begin{tikzpicture}

\begin{axis}[%
width=4in,
height=3in,
scale only axis,
clip=false,
xmin=0.1,
xmax=1,
xlabel style={font=\color{white!15!black}},
xlabel={\Large NMIL},
ymin=0,
ymax=2.5,
ylabel style={font=\color{white!20!black}},
ylabel={\Large $ \max_{y \in \Y} \hspace{2pt} \omega(y) $},
axis background/.style={fill=white},
xmajorgrids,
ymajorgrids,
legend style={legend cell align=left, align=left, draw=white!15!black}
]
\addplot [color=blue, line width=1.2pt, mark=o, mark options={solid, blue},  mark size=3.0pt]
  table[row sep=crcr]{%
0.614055979843306	0.249989290436101\\
0.40749218006555	0.467312806963884\\
0.299882004727968	0.694627729962609\\
0.252336877359735	0.898729525285485\\
0.222001506768911	1.13200841745822\\
0.195460632929228	1.38806348769694\\
0.169494509817045	1.64362547965089\\
0.144196061764307	1.8876744429263\\
0.121211717113631	2.12990401298736\\
0.100839132874635	2.36768416964981\\
};
\addlegendentry{Algortihm 1}

\node[right, align=left]
at (axis cs:0.62,0.271) {$\text{ }\epsilon\text{=0.25}$};
\node[right, align=left]
at (axis cs:0.41,0.539) {$\text{ }\epsilon\text{=0.5}$};
\node[right, align=left]
at (axis cs:0.305,0.766) {$\text{ }\epsilon\text{=0.75}$};
\node[right, align=left]
at (axis cs:0.259,0.932) {$\text{ }\epsilon\text{=1}$};
\node[right, align=left]
at (axis cs:0.229,1.166) {$\text{ }\epsilon\text{=1.25}$};
\node[right, align=left]
at (axis cs:0.201,1.412) {$\text{ }\epsilon\text{=1.5}$};
\node[right, align=left]
at (axis cs:0.176,1.672) {$\text{ }\epsilon\text{=1.75}$};
\node[right, align=left]
at (axis cs:0.152,1.921) {$\text{ }\epsilon\text{=2}$};
\node[right, align=left]
at (axis cs:0.129,2.159) {$\text{ }\epsilon\text{=2.25}$};
\node[right, align=left]
at (axis cs:0.111,2.378) {$\text{ }\epsilon\text{=2.5}$};
\addplot [color=red, line width=1.2pt, mark=square, mark options={solid, red},  mark size=3.0pt]
  table[row sep=crcr]{%
1	1.43773881688958e-16\\
0.997978869972794	0.0109117946383655\\
0.9789580229695	0.156629674501263\\
0.918817040609376	0.593802825976573\\
0.821946801040326	1.03730121310417\\
0.70211502622707	1.36512504239354\\
0.578666240060208	1.63524406964528\\
0.460278819984222	1.88190863864017\\
0.356422013842445	2.12646342494424\\
0.267552114870886	2.36591508826868\\
};
\addlegendentry{Complete merging}

\node[right, align=left]
at (axis cs:1.006,0.005) {$\text{ }\epsilon\text{=0.25}$};
\node[right, align=left]
at (axis cs:1.01,0.104) {$\text{ }\epsilon\text{=0.5}$};
\node[right, align=left]
at (axis cs:0.85,0.16) {$\text{ }\epsilon\text{=0.75}$};
\node[right, align=left]
at (axis cs:0.925,0.627) {$\text{ }\epsilon\text{=1}$};
\node[right, align=left]
at (axis cs:0.825,1.085) {$\text{ }\epsilon\text{=1.25}$};
\node[right, align=left]
at (axis cs:0.704,1.422) {$\text{ }\epsilon\text{=1.5}$};
\node[right, align=left]
at (axis cs:0.577,1.707) {$\text{ }\epsilon\text{=1.75}$};
\node[right, align=left]
at (axis cs:0.463,1.949) {$\text{ }\epsilon\text{=2}$};
\node[right, align=left]
at (axis cs:0.364,2.169) {$\text{ }\epsilon\text{=2.25}$};
\node[right, align=left]
at (axis cs:0.272,2.404) {$\text{ }\epsilon\text{=2.5}$};
\end{axis}

\end{tikzpicture}%